\documentclass[conference,10pt]{IEEEtran}
\IEEEoverridecommandlockouts
\usepackage{amsmath,amssymb,amsfonts}
\usepackage{algorithmic}
\usepackage{graphicx}
\usepackage{textcomp}
\usepackage{xcolor}
\def\BibTeX{{\rm B\kern-.05em{\sc i\kern-.025em b}\kern-.08em
    T\kern-.1667em\lower.7ex\hbox{E}\kern-.125emX}}

\usepackage{bm,marvosym,mathtools,environ,array,amsthm,url}
\usepackage[ruled,vlined,linesnumbered,resetcount]{algorithm2e}

\graphicspath{{./figs/}}

\renewcommand{\v}[1]{\boldsymbol{\mathbf{#1}}} 
\newcommand{\opn}[1]{\operatorname{#1}} 
\newcommand{\supp}[1]{\operatorname{supp}\left(\v{#1}\right)} 
\newcommand{\card}[1]{\left|\mathbb{#1}\right|} 
\newcommand{\set}[1]{\mathbb{#1}} 
\newcommand{\norm}[2]{\lVert #1 \rVert_{#2}} 
\newcommand{\braces}[1]{\left\{#1\right\}} 
\newcommand{\tm}[0]{\times} 
\newcommand{\prt}[1]{\left(#1\right)} 
\newcommand{\sqb}[1]{\left[#1\right]} 

\newcommand{\inc}[1]{\in \set{C}^{#1}} 
\newcommand{\inr}[1]{\in \set{R}^{#1}} 

\newtheorem{theorem}{Theorem}
\newtheorem{definition}{Definition}

\makeatletter
\newsavebox\myboxA
\newsavebox\myboxB
\newlength\mylenA
\newcommand*\xov[2][0.75]{%
    \sbox{\myboxA}{$\m@th#2$}%
    \setbox\myboxB\null
    \ht\myboxB=\ht\myboxA%
    \dp\myboxB=\dp\myboxA%
    \wd\myboxB=#1\wd\myboxA
    \sbox\myboxB{$\m@th\overline{\copy\myboxB}$}
    \setlength\mylenA{\the\wd\myboxA}
    \addtolength\mylenA{-\the\wd\myboxB}%
    \ifdim\wd\myboxB<\wd\myboxA%
       \rlap{\hskip 0.5\mylenA\usebox\myboxB}{\usebox\myboxA}%
    \else
        \hskip -0.5\mylenA\rlap{\usebox\myboxA}{\hskip 0.5\mylenA\usebox\myboxB}%
    \fi}
\makeatother

\NewEnviron{meq}{
    \begin{align}
        \begin{split}
            \BODY
        \end{split}
    \end{align}
    }

\begin{document}

\title{\huge Study of Noncoherent Sparse Subarrays for Direction Finding Based on Low-Rank and Sparse Recovery\\

\thanks{Funding: Pontifical Catholic University of Rio de Janeiro, FAPERJ and CNPq (Brazil).}
}

\author{\IEEEauthorblockN{Wesley S. Leite}
\IEEEauthorblockA{\textit{Center for Telecommunications Research (CETUC)} \\
\textit{Pontifical Catholic University of Rio de Janeiro (PUC-Rio)}\\
Rio de Janeiro, Brazil \\
wleite@ieee.org}
\and
\IEEEauthorblockN{Rodrigo C. De Lamare}
\IEEEauthorblockA{\textit{Center for Telecommunications Research (CETUC)} \\
\textit{Pontifical Catholic University of Rio de Janeiro (PUC-Rio)}\\
Rio de Janeiro, RJ\\
delamare@puc-rio.br}

}

\maketitle

\begin{abstract}
This paper investigates the problem of noncoherent direction-of-arrival (DOA) estimation using different sparse subarrays. In particular, we present a Multiple Measurements Vector (MMV) model for noncoherent DOA estimation based on a low-rank and sparse recovery optimization problem. Moreover, we develop two different practical strategies to obtain sparse arrays and subarrays: i) the subarrays are generated from a main sparse array geometry (Type-I sparse array), and ii) the sparse subarrays that are directly designed and grouped together to generate the whole sparse array (Type-II sparse array). Numerical results demonstrate that the proposed MMV model can benefit from multiple data records and that Type-II sparse noncoherent arrays are superior in performance for DOA estimation.
\end{abstract}


\section{Introduction}
In the last decade, the fields of sparse reconstruction, Compressive Sensing (CS) and direction-of-arrival (DOA) estimation with arrays of sensors have been largely investigated in applications involving sonar, radar and communications \cite{Malioutov2005,VanTrees2002,Zhang2013,Pal2010,jiodoa,mskaesprit}. In this sense, a great deal of research has been devoted to sparse sensor arrays and sparse signal processing techniques due to their remarkable performance improvements in beamforming applications, as well as direction-of-arrival (DOA) estimation \cite{Liu2016,Liu2016-2,Leite2021,Pal2010-1,Pal2012,jiodoa,mskaesprit,jioalrd,Qiu2016,rdstapcp,misc,emisc}.

In this sense, many different signal processing DOA estimation techniques resort to sparse signal processing and subspace strategies \cite{jidf,sjidf,aifir,rrber,jio,jiocdma,jiomimo,jiostap,jiobf,wlmwf,wljio,wlbd,l1stap,ccg,sacg,sastap,rdrcb,saalt,sacg,locsme,okspme,lrcc}, since they present some advantages over standard approaches. Indeed, they are able to recover the DOAs even for an incredibly small amount of data records (even a single snaphot) \cite{Fortunati2014}. CS algorithms like Orthogonal Matching Pursuit (OMP) and Iterative Hard Thresholding (IHT) have been widely used to solve this problem \cite{Pati1993,Blumensath2009,dce}.

In addition to sparse models, the large development of sparse linear arrays (SLA) or non-uniform arrays, have also drawn interest of recent cutting-edge research, since they are capable of recovering more DOAs than the number of physical sensors with some additional processing (difference coarray, for example). More recently, the advantages of SLAs have been extended to subarrays applications \cite{Gu2011,Elbir2019,See2004,Swindlehurst2001,Rieken2004}.

In this paper we investigate the problem of noncoherent direction-of-arrival (DOA) estimation using sparse subarrays with distinct local oscillators. In particular, we present a Multiple Measurements Vector (MMV) model for noncoherent DOA estimation based on a low-rank and sparse recovery optimization problem. Moreover, we develop two different practical strategies to obtain sparse arrays and subarrays: i) the subarrays are generated from a main sparse array geometry (Type-I sparse array), and ii) the sparse subarrays that are directly designed and grouped together to generate the whole sparse array (Type-II sparse array). Numerical results demonstrate that the proposed MMV model can benefit from multiple data records and that Type-II sparse noncoherent arrays are superior in performance for DOA estimation.

\emph{Paper structure}: In Section~\ref{sec:systemModel}, the standard single snapshot data model is presented. In Section~\ref{sec:propMethod}, the multiple snapshot model is derived and a sparse recovery strategy is proposed. Moreover, the idea of Type-I and Type-II arrays is introduced. Section~\ref{sec:analysis} provides an analysis of the manifold structure and the degrees of freedom. In Section~\ref{sec:simulation}, the numerical simulations demonstrating both approaches are exhibited and discussed, whereas Section~\ref{sec:conclusion} draws the conclusions.

\emph{Notation}: $\set{S}$, $a$, $\v{a}$ and $\v{A}$ indicate sets, scalars, column vectors, and matrices, respectively. $\card{\set{S}}$ means the cardinality of set $\set{S}$ (number of elements). $\opn{blkdiag}(\cdot)$ and $\opn{supp}(\cdot)$ are the block diagonal matrix and support set operators, respectively, whereas $\opn{vecd}(\cdot)$ is the diagonal extraction operator and $\opn{vec}_{a\tm b}^{-1}(\cdot)$ is an operator that transforms a column vector into a matrix with dimension $a\tm b$ by stacking each $a$ of its elements into a different column sequentially. $\opn{H}_D(\cdot)$ is the hard thresholding operator.

\section{System Model and Problem Statement}\label{sec:systemModel}

The representation of the data acquisition model with noncoherent subarrays for a single snapshot (Single Measurement Vector - SMV) is given by \cite{Tirer2021}
\begin{equation}\label{eq:smv_model_dense}
    \v{x}^{(l)} = e^{-j\phi_l}\v{A}_{\set{S}_l}(\v{\theta})\v{s}+\v{n}_{\set{S}_l}\text{, }l=1,\ldots,L
\end{equation}
where $\phi_l$ is the $l$-th subarray phase shift, $\v{A}_{\set{S}_l}(\theta)\inc{N_l\tm D}$ is the $l$-th subarray manifold with geometry defined by the set of integers $\set{S}_l$ (normalized positions in terms of integer multiples of $d$ - minimum intersensor spacing), the $l$-th subarray has $N_l$ sensors, and there are $D$ impinging sources with normalized directions given by $\v{\theta}\in [-1,1)^{D}$ (sine of DOAs - spatial frequency), $\v{s}\inc{D}$ is the source signal, $\v{x}^{(l)}\inc{N_l}$ is the $l$-th subarray received signal, and $\v{n}_{\set{S}_l}$ is the subarray noise vector. The noise and the source signal are drawn from a zero-mean circularly complex multivariate Gaussian distribution. The noise is white and the sources are uncorrelated. Remark: as abuse of notation, we refer to the set $\set{S}$ defining the sensors locations as the array itself.

If we consider a sparse reconstruction model, i.e., the DOAs are replaced by a search grid with $g$ points, the model in (\ref{eq:smv_model_dense}) can be recast in matrix form as
\begin{equation}\label{eq:smv_model}
    \v{x} = \v{\Gamma}(\v{\phi})\v{A}_\set{S}(\tilde{\v{\theta}})\tilde{\v{s}}+\v{n}_{\set{S}}
\end{equation}
where $\v{x}\inc{N}$ is the array received signal $\prt{N=|\set{S}|=\sum_{l=1}^{L}N_l}$, $\tilde{\v{\theta}}\in [-1,1)^{g}$ is the grid search vector which contains the true normalized DOAs, $\set{S}=\bigcup_{l=1}^{L}\set{S}_l$ is the set of integers that define the whole array geometry, $\v{\Gamma}(\v{\phi})\inc{N\times N}$ is a matrix that accounts for the multiple phase shifts for each noncoherent subarray such that $\v{\phi}\in [0,2\pi)^{L}$, $\v{A}_\set{S}(\tilde{\v{\theta}}) \inc{N\times g}$ is the overcomplete array manifold, $\tilde{\v{s}}\inc{g}$ is the sparse source vector with support set cardinality given by $|\opn{supp}(\tilde{\v{s}})|=D$, and $\v{n}_{\set{S}}\inc{N}$ is the noise vector. The phase shifts matrix $\v{\Gamma}(\v{\phi})$ is defined as
\begin{equation}
    \v{\Gamma}(\v{\phi}) = \opn{diag}\prt{\alpha_1^{\ast}\v{I}_{N_1},\ldots,\alpha_L^{\ast}\v{I}_{N_L}}
\end{equation}
where $\alpha_l=e^{j\phi_l}$ is the phase shift associated to each of the $L$ noncoherent subarrays. Notice also that the array received signal and manifold have the following matrix block representation,
\begin{equation}
    \v{x} = \sqb{(\v{x}^{(1)})^T,\ldots,(\v{x}^{(L)})^T}^T
\end{equation}
\begin{equation}
    \v{A}_\set{S}(\tilde{\v{\theta}}) = \sqb{\v{A}_{\set{S}_1}^T(\tilde{\v{\theta}}), \ldots, \v{A}_{\set{S}_L}^T(\tilde{\v{\theta}})}^T
\end{equation}
Note that the set notation is extensively used in this paper to emphasize the dependence of the equations from the subarray geometries $\set{S}_l$. The problem that we would like to solve is to find the support set of $\tilde{\v{s}}$, which directly determines the DOAs from the grid $\tilde{\v{\theta}}$, where both $\tilde{\v{s}}$ and $\v{\phi}$ are unknowns.

\section{Proposed DOA Estimation Method}\label{sec:propMethod}

In this section, we extend the convex optimization method based on low rank and sparse recovery devised in \cite{Tirer2021} to a multiple snapshots scenario considering two approaches to generate the subarray geometries. In this sense, we employ a multiple snapshots scenario with a double sparse signal processing approach: sparse model and sparse array. To this end, we introduce Type-I and Type-II Sparse Linear Arrays.

\subsection{Sparse and low-rank recovery with multiple snapshots}

The extended version of the SMV model in Eq. (\ref{eq:smv_model}) for $T$ snapshots can be written in matrix form as the following MMV model
\begin{equation}\label{eq:mmv_model}
    \xov{\v{X}} = \xov{\v{\Gamma}}\xov{\v{A}}\xov{\v{S}}+\xov{\v{N}}
\end{equation}
where
\begin{equation}
    \xov[1]{\v{\Gamma}} = \sqb{\v{\Gamma}(\v{\phi}_1)|\ldots|\v{\Gamma}(\v{\phi}_T)}\inc{N\times NT}
\end{equation}
\begin{equation}
\xov{\v{A}} = \opn{blkdiag}\prt{\v{A},\ldots,\v{A}}\inc{NT\times gT}
\end{equation}
\begin{equation}
\xov{\v{S}} = \opn{blkdiag}\prt{\tilde{\v{s}}_1,\ldots,\tilde{\v{s}}_T}\inc{gT\times T}
\end{equation}
\begin{equation}
\xov{\v{N}} = [\v{n}_1, \ldots, \v{n}_{\text{T}}]\inc{N\times T}
\end{equation}
\begin{equation}
    \xov{\v{X}} = [\v{x}_1,\ldots,\v{x}_{\text{T}}]\inc{N\times T}
\end{equation}
where $\v{A}=\v{A}_\set{S}(\tilde{\v{\theta}})$.

Recall that we have incorporated the additional constraint that the phase shifts change from snapshot to snapshot and between the subarrays. Thus, one would have $LT$ different phase shifts for $L$ arrays. Notice also that the model in (\ref{eq:mmv_model}) is a general case for the partially calibrated array scenario. Indeed, it suffices to assume $\v{\Gamma}(\v{\phi}_t) = \v{\Gamma}(\v{\phi}),~\forall t=1,\ldots,T$. Moreover, since the matrices $\xov{\v{\Gamma}}$, $\xov{\v{A}}$, and $\xov{\v{S}}$ are sparse, additional signal processing techniques can be employed to reduce the total cost when adopting the model.

In order to solve the SMV model in (\ref{eq:smv_model}), the authors in \cite{Tirer2021} proposed a convex optimization formulation derived from a bilinear arrangement and a convex relaxation procedure. Their algorithm is based on the solution of the following problem
\begin{equation}\label{eq:cvx_prob}
    \begin{array}{r@{}l}
        \textrm{minimize}&{}~~~\norm{\v{G}}{1,2}+\varepsilon\norm{\v{G}}{\ast}\\
        \textrm{subject to}&{}~~~\sum_{l=1}^{L}\norm{\v{x}^{(l)}-\v{A}_{\set{S}_l}(\tilde{\v{\theta}})\v{g}_l}{2}^2\leq CM\sigma^2\\
    \end{array}
\end{equation}
where $\v{G}=[\v{g}_1,\ldots,\v{g}_L]\inc{g\tm L}$ is a row-sparse optimization variable. This problem can be effectively solved using the software package CVX \cite{cvx}. Moreover, $\v{G}=\tilde{\v{s}}\v{\alpha}^H$ (rank-one), which allows us to perform an SVD and recover a proxy to $\tilde{\v{s}}$ as $\hat{\v{s}}=\sigma_i\hat{\v{u}}_i$, where $\hat{\v{u}}_i$ is the right-singular vector corresponding to the largest singular value. Thus, the support $\supp{\hat{\v{s}}}$ determines the true directions from the search grid $\tilde{\v{\theta}}$. Indeed, since $\opn{supp}(\hat{\v{u}}_i)=\opn{supp}(\hat{\v{s}})$, we propose to resort only to $\hat{\v{u}}_i$. Inspired by the work in \cite{Malioutov2005}, we recover each of the proxies to the support sets of the solutions $\hat{\v{s}}_t$ according to the following:
\begin{equation}
    \xov{\v{U}} = \opn{blkdiag}\prt{\hat{\v{u}}_i^{(1)},\ldots,\hat{\v{u}}_i^{(T)}}
\end{equation}
After that, we collapse the columns and perform an inverse vectorization operation such that $\v{E}=\opn{vec}_{g\tm T}^{-1}\prt{\xov{\v{U}}\v{1}_{T}}\inc{g\tm T}$,
which is ideally row-sparse. From that, we obtain the squared energy associated to the spatial index for each row of $\v{E}$ and recover its pseudo-spectrum through $\opn{vecd}(\v{E}\v{E}^H)$. The peaks of the pseudo-spectrum should be used to recover the estimated DOAs from the search grid. The algorithm is summarized in Algorithm~\ref{alg:noncoh_sparserec}.

While this procedure can be considered costly and having the disadvantage of not taking into account the synergy between the snapshots, we numerically verified that the algorithm performs reasonably well when the data record is small to moderate. However, if large amounts of data are available, one could adapt the algorithm to incorporate the SOCP $l_1$-SVD technique for the multiple snapshots model \cite{Malioutov2005}. On the other hand, we also point out that breaking the original array into subarrays has the potential of reducing the computational burden associated to many signal processing operations.

\begin{algorithm}[ht]
	\DontPrintSemicolon
	\SetKwInOut{Inp}{Input}
	\SetKwInOut{Out}{Output}

	\Inp{Overcomplete array manifold $\v{A}(\tilde{\v{\theta}})$, observed data $\xov{\v{X}}$, number of sensors for each subarray $N_l$, grid $\v{\theta}^g\inr{g}$, number of sources $D$}

	\BlankLine
	$\set{T}^{(0)}=\braces{g+1}$, $\v{h}^{(0)}=\v{0}$\label{alg:lbml-omp_initialize}
	\BlankLine

	\For{$t\leftarrow 1$ \KwTo $T$}{
	solve (\ref{eq:cvx_prob}) with $\v{x}_t$\;
	perform economy-size SVD on $\hat{\v{G}}$ and obtain $\hat{\v{u}}_i^{(t)}$\;
	}
	Calculate $\v{E}=\opn{vec}_{g\tm T}^{-1}\prt{\xov{\v{U}}\v{1}_{T}}$\;
	\tcp{Computes the grid support set}
	$\set{T} = \opn{supp}\prt{\opn{H}_D\prt{\opn{vecd}\prt{\v{E}\v{E}^{H}}}}$\;
	\BlankLine
	\caption{Multiple Snapshot Non-Coherent Low-Rank and Sparse Recovery Algorithm}
	\Out{Estimated DOAs $\hat{\v{\theta}} = \tilde{\v{\theta}}_{\set{T}}$}
	\label{alg:noncoh_sparserec}
\end{algorithm}

\subsection{Type-I and Type-II array geometries}

In this section, we introduce two approaches to tackle the problem of noncoherent processing regarding the subarray geometries. For the trivial case, we have the ULA geometry. Since this geometry is uniform, if we split it into linear adjacent segments of arbitrary sizes, these segments are still ULAs with a smaller number of sensors. However, for sparse arrays, this is not necessarily true. In what follows, we systematize the definitions of Type-I and Type-II Sparse Linear Arrays (Type-I SLA and Type-II SLA).

\begin{definition}[Type-I Sparse Linear Array]
    A Type-I Sparse Linear Array (Type-I SLA) corresponds to an array of predefined sparse geometry $\set{S}$. The subsets defining the subarray geometries are generated from partitions $\set{S}_l\subset\set{S}$ such that if $s\in \set{S}_i$ and $f\in \set{S}_j$, $i<j$, then $s<f$.
\end{definition}

\begin{definition}[Type-II Sparse Linear Array]\label{def:IISLA}
    A Type-II Sparse Linear Array (Type-II SLA) corresponds to a union of subarrays with predefined sparse linear geometries $\set{S}_l$ (partitions of the array geometry $\set{S}$). The set $\set{S}_1$ defines the reference subarray. The remaining subarrays are derived from $\set{S}_1$ through $\set{S}_i=\set{S}_{i-1}^{\Delta_{i-1}}$, where $\Delta_{i-1}$ is a translation factor for all the elements of $\set{S}_{i-1}$ and is given by $\Delta_{i-1}=\mu+\kappa_{i-1}$. $\mu$ is the normalized distance between subarrays (in terms of integer multiples of $d$) and $\kappa_{i-1}$ is the aperture of the $(i-1)$-th subarray.
\end{definition}
Notice that the definitions in both cases restrict the subarrays such that they attend to the following rules: i) they do not share any sensors (partitions are pairwise disjoint); ii) they are collinear; and iii) the array is given by the union of the subarrays $\prt{\set{S}=\bigcup_{l=1}^L \set{S}_l}$. Additionally, observe that if we apply the additional constraint  $N_l=N/L$ (or $\kappa$ is a constant for all the subarrays), $\forall l\in\{1,\ldots,L\}$, then the subarrays derived from Type-II SLA become multiple invariant \cite{Swindlehurst2001}.

The key difference between both sparse array definitions is that for Type-I SLA, the subarrays are generated from a splitting of the array with a predefined sparse geometry. On the other hand, for Type-II SLA, the array is generated by the union of sparse linear subarrays with a predefined geometry.

To illustrate, consider the sparse geometry defined according to the set
$\set{S}^{\text{MRA}} = \{0,1,3,6,13,20,27,31,35,36\}$, which corresponds to a Minimum Redundancy Array with $N=10$ sensors \cite{Moffet1968}. We adopt $L=2$ subarrays. The corresponding Type-I array is the original array itself $\prt{\set{S}^{\text{I-MRA}}=\set{S}^{\text{MRA}}}$ and its subarrays are given by $\set{S}_{1}^{\text{I-MRA}}=\{0,1,3,6,13\}$ and $\set{S}_{2}^{\text{I-MRA}}=\{20,27,31,35,36\}$. On the other hand, consider two MRAs with $N=5$ sensors each given by $\set{S}_{1}^{\text{II-MRA}}=\{0,1,4,7,9\}$ and $\set{S}_{2}^{\text{II-MRA}}=\set{S}_{1}^{\Delta_1;\text{II-MRA}}=\{10,11,14,17,19\}$, with translation factor $\Delta_1=1+(9-0)=10$. Thus, $\set{S}^{\text{MRA}}=\set{S}^{\text{I-MRA}}\neq \set{S}^{\text{II-MRA}}$.

From this, we conclude that signal processing techniques can be employed in real time to split the arrays conveniently (Type-I case). However, if we intend to keep all the already well-established good properties for some geometries in generating the subarrays, we should pay attention to the fact that the arrays should be defined a priori (Type-II). In our simulations, we empirically have shown that Type-II SLAs are way better than Type-I when estimating the DOAs from a noncoherent processing perspective.

\section{Analysis of manifold structure and degrees of freedom}\label{sec:analysis}

A natural question that arises from the above introduced (sub)array geometries is related to the manifold structure, as well as the degrees of freedom (DoF) for a difference coarray scenario \cite{Leite2021,Liu2016} (although this coarray structure is not exploited in this paper). In this section, we shed some light upon these two important aspects.

\subsubsection{Type-I}
Clearly, there is no \emph{a priori} analytical relation between the subarray manifolds, due to the fact that the geometries change dramatically between subarrays. Also, the number of DoF for the predefined geometry and its Type-I counterpart is obviously the same, since both geometries coincide. Moreover, there is no general rule to predict the DoF associated to the subarrays. This requires an analysis case by case for the sparse geometry that is chosen to be employed.

\subsubsection{Type-II}
Some analytical relations can be straightforwardly derived. In this case, to simplify the equations, we will assume that the subarrays are multiple invariant, i.e., they attend the sufficient condition of having the same number of physical sensors $N/L$ and thus the same aperture. Using the array manifold of the whole array expressed in matrix form, we can write
\begin{equation}\label{eq:type_II_manif}
    \v{A}_{\set{S}}=\sqb{\v{A}_{\set{S}_1}^T,\prt{\v{A}_{\set{S}_1}\v{\Lambda}^{\Delta}}^T,\ldots,\prt{\v{A}_{\set{S}_1}\v{\Lambda}^{(L-1)\Delta}}^T}^T
\end{equation}
where we dropped the dependence of the manifolds on the DOAs to simplify the notation. The matrix $\v{\Lambda}$ is defined as
\begin{equation}
    \v{\Lambda} = \opn{diag}\prt{e^{j\pi\theta_1},\ldots,e^{j\pi\theta_D}}
\end{equation}
The multiple exponents of $\v{\Lambda}$ in (\ref{eq:type_II_manif}) represent the subarrays translations along the straight line and are assumed to be $\Delta=\mu+\kappa$ (see Definition~\ref{def:IISLA}).

In what follows, we establish that the number of DoF for Type-II arrays is upper-bounded by a function of the DoF of the subarrays (sDoF), which can be theoretically calculated for a variety of preconceived geometries.

\begin{theorem}
Consider a Type-II array with geometry as defined in Definition~\ref{def:IISLA} with equal-aperture subarrays. If $1\leq \mu \leq \kappa$, then the number of \emph{DoF} of the array $\set{S}$ is upper-bounded by \emph{$L(\text{sDoF}-1)+2(L-1)\mu+1$}, where \emph{sDoF} is the number of \emph{DoF} for each subarray. If $\mu>\kappa$, then the number of \emph{DoF} is equal to \emph{$(2L-1)\text{sDoF}$}.
\proof Let $c(n)$ be a discrete-valued function that assumes the value of 1 if there is a sensor at $nd$ or 0 otherwise. This function for a Type-II array is given by $c(n)=c_1(n)+\ldots+c_{L}(n)$. Since the subarrays are translated versions of the reference array, $c_l(n)=c_1(n-\Delta_l)$, where $\Delta_l=(l-1)\Delta$. The weight function associated to that array (counts the number of spatial correlations with lag $n$ is defined through \cite{Pal2010-1} $w(n) = c(n)\circledast c^{-}(n)$, where $c^{-}(n)$ is the time reversal version of $c(n)$. Then, one can write
\begin{meq}\label{eq:proof_theo}
    w(n) & = \prt{\sum_{i=1}^L c_i(n)} \circledast \prt{\sum_{j=1}^L c_j^{-}(n)}\\ 
         & = \sum_{i,j=1}^L c_1(n-(i-1)\Delta)\circledast c_{1}^{-}(n-(j-1)\Delta)\\
         & = \sum_{i,j=1}^L c_1(n)\circledast c_{1}^{-}(n)\circledast\delta(n-\Delta_i+\Delta_j)\\
         & = \sum_{i,j=1}^L w_1(n-\Delta_i+\Delta_j)\\
\end{meq}
The number of DoF is the cardinality of the support of $w(n)$. This support set has always an odd number of elements, given that $w(n)$ is an even function. Clearly, from (\ref{eq:proof_theo}), the weight function of the reference array $w_1(n)$ is repeated along the domain with displacement factors given by $\Delta_j-\Delta_i$. Since $|\opn{supp}(w_1(n))|=\text{sDoF}$ and for $\mu>\kappa$ there is no superposition of the weight functions of the subarrays with a different $\Delta_j-\Delta_i$, then $|\opn{supp}(w(n))|=(2L-1)|\opn{supp}(w_1(n))|\Rightarrow \text{DoF}=(2L-1)\text{sDoF}$. Note that there are $2L-1$ diagonals in a $L\tm L$ matrix representing all the possible $\Delta_j-\Delta_i$. For $1\leq \mu\leq \kappa$, the superposition implies that the support set of $w(n)$ ranges from $-(L-1)\Delta-\kappa$ to $(L-1)\Delta+\kappa$. Since $\kappa=(\text{sDoF-1})/2$, the support set of $w(n)$ has a maximum number of elements equal to $2\sqb{(L-1)\Delta+\kappa}+1$ or $L(\text{sDoF}-1)+2(L-1)\mu+1$. The equality holds if the difference coarray of the subarrays has no holes (no missing lags), as it indeed happens for many geometries like Two-Level Nested and (restricted) Minimum Redundancy Arrays. \qed
\end{theorem}

\section{Simulation}\label{sec:simulation}

In what follows, we illustrate with computer simulations the algorithm performance under different scenarios employing Type-I and Type-II sparse arrays. For that, we employ Uniform Linear Arrays (ULA), Two-level Nested Arrays (NAQ2), and (restricted) Minimum Redundancy Arrays (MRA) \cite{VanTrees2002,Pal2010-1,Moffet1968,Tuncer:2009:CMD:1795749}. Notice that the MRAs considered in this paper are not zero-redundancy. Instead, they allow some repetition of spatial correlations, but presents the largest hole-free difference coarray (restricted MRA) for a given number of physical sensors.

The simulation scenarios adopt $L=2$ subarrays with $N=6$ sensors each, $d/\lambda=1/2$, and $\mu=1$. The performance curves are drawn by assessing the Root Mean Square Error (RMSE) \cite{VanTrees2002} $\textrm{RMSE}=\sqrt{\frac{1}{DR}\sum_{i=1}^{R}\norm{\v{\theta}-\hat{\v{\theta}}_i}{2}^2}$.
We also add that we use $R=500$ Monte Carlo runs in order to have well behaved curves and the phase shifts $\v{\phi}_t$ are drawn from $U(0,2\pi)$ (uniform distribution) for each subarray, snapshot and run. The sources DOAs are considered to be coming from the following normalized directions: $\v{\theta}=\sqb{0,0.2,0.4,0.6,0.8}$. The amount of data for the SNR curves is $T=10$ snapshots and $\text{SNR}=10$ dB for the curves against the snapshots.

\begin{figure}[htbp]
\centering
\includegraphics[width=.47\textwidth]{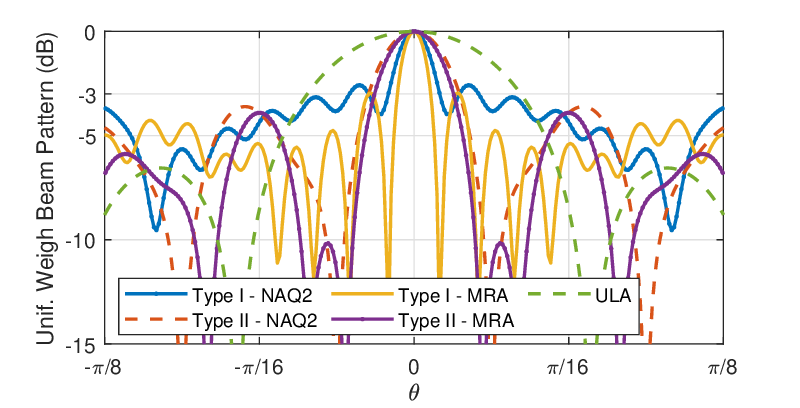}
\caption{Beampattern for Type-I and Type-II arrays.}
\label{fig:0}
\end{figure}

Firstly, we describe the two types of sparse arrays and analyze their uniformly weighted beampattern. Fig.~\ref{fig:0} shows a comparison between the beampattern for Types I and II SLA for multiple geometries, as well as the ULA. It can be clearly seen that both sparse structures exhibit a very particular response and Type-I SLAs have narrower main lobes. Although this is a desirable feature to increase source discrimination capabilities, notice that their sidelobes are not as low as their Type-II counterparts. We also add that the ULA has the worst beampattern mainlobe and sidelobe responses. Then, in general we expect its performance to have some important degradation in comparison to both types of sparse arrays.

\begin{figure}[htbp]
\centering
\includegraphics[width=.47\textwidth]{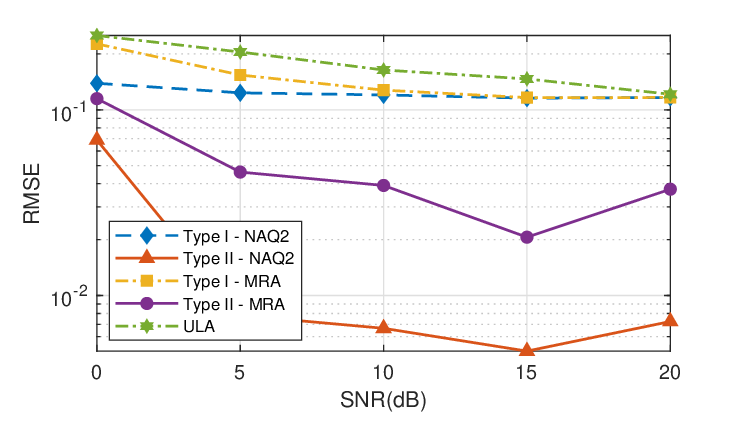}
\caption{RMSE performance curves against SNR for Type-I and Type-II arrays. $T=10$ snapshots. $D=5$ sources located at $\theta=[0,0.2,0.4,0.6,0.8]$.}
\label{fig:1}
\end{figure}

Fig.~\ref{fig:1} shows the RMSE against the SNR. Clearly, Type-II SLAs present a much better performance in comparison to Type-I. Moreover, it is clear that the sparse arrays also present an increased performance in comparison to the ULA geometry. In our experiments, we have verified that this gap in performance becomes more prominent as we increase the number of snapshots. This can be explained by the larger aperture that the sparse arrays have in comparison to uniform arrays, whereas the increased performance of Type-II over Type-I can be explained by the fact that the former preserves all the good properties from the original array design.

\begin{figure}[htbp]
\centering
\includegraphics[width=.47\textwidth]{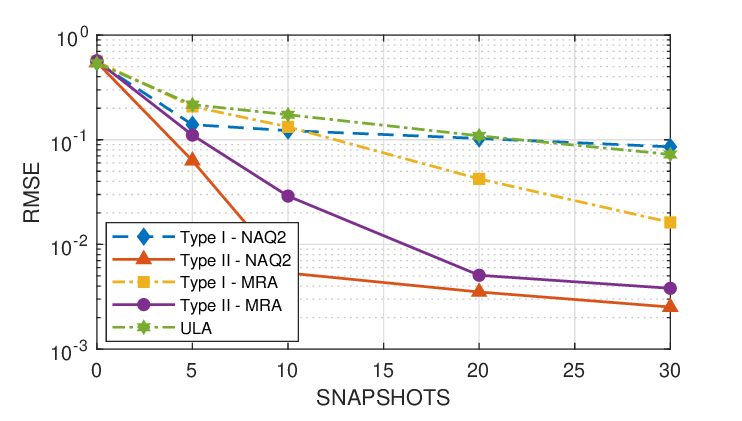}
\caption{RMSE performance curves against snapshots for Type-I and Type-II arrays. $\text{SNR}=10$ dB. $D=5$ sources located at $\theta=[0,0.2,0.4,0.6,0.8]$.}
\label{fig:2}
\end{figure}

Fig.~\ref{fig:2} shows a comparison between the different arrays against the snapshots. It is clear that the benefits of sparse arrays for noncoherent processing increase as we use larger amounts of data. Again, the Type-II arrays have better performance when compared to Type-I.

\section{Conclusion}\label{sec:conclusion}
In this paper, we have discussed noncoherent sparse arrays and their use to perform DOA estimation. We have devised an extension of a joint low-rank and sparse formulation technique to account for multiple snapshots in DOA estimation. Moreover, we introduced the idea of generating the sparse arrays through the splitting of a larger well-known sparse array (Type-I) or through the gathering of sparse subarrays with a predefined geometry (Type-II). Particularly, we have come up with numerical evidence that Type-II arrays present better performance for DOA estimation with noncoherent processing.

\bibliographystyle{IEEEtran}
\bibliography{mybib}

\begin{thebibliography}{10}
\providecommand{\url}[1]{#1}
\csname url@samestyle\endcsname
\providecommand{\newblock}{\relax}
\providecommand{\bibinfo}[2]{#2}
\providecommand{\BIBentrySTDinterwordspacing}{\spaceskip=0pt\relax}
\providecommand{\BIBentryALTinterwordstretchfactor}{4}
\providecommand{\BIBentryALTinterwordspacing}{\spaceskip=\fontdimen2\font plus
\BIBentryALTinterwordstretchfactor\fontdimen3\font minus \fontdimen4\font\relax}
\providecommand{\BIBforeignlanguage}[2]{{%
\expandafter\ifx\csname l@#1\endcsname\relax
\typeout{** WARNING: IEEEtran.bst: No hyphenation pattern has been}%
\typeout{** loaded for the language `#1'. Using the pattern for}%
\typeout{** the default language instead.}%
\else
\language=\csname l@#1\endcsname
\fi
#2}}
\providecommand{\BIBdecl}{\relax}
\BIBdecl

\bibitem{Malioutov2005}
D.~Malioutov, M.~{\c{C}}etin, and A.~S. Willsky, ``{A sparse signal reconstruction perspective for source localization with sensor arrays},'' \emph{IEEE Transactions on Signal Processing}, vol.~53, no.~8, pp. 3010--3022, 2005.

\bibitem{VanTrees2002}
H.~L. {Van Trees}, \emph{Optimum Array Processing: Part IV of Detection, Estimation, and Modulation Theory}.\hskip 1em plus 0.5em minus 0.4em\relax New York, USA: John Wiley {\&} Sons, Inc., mar 2002.

\bibitem{Zhang2013}
Y.~D. Zhang, M.~G. Amin, and B.~Himed, ``{Sparsity-based DOA estimation using co-prime arrays},'' \emph{ICASSP, IEEE International Conference on Acoustics, Speech and Signal Processing - Proceedings}, pp. 3967--3971, 2013.

\bibitem{Pal2010}
P.~Pal and P.~P. Vaidyanathan, ``{A novel array structure for directions-of-arrival estimation with increased degrees of freedom},'' in \emph{2010 IEEE International Conference on Acoustics, Speech and Signal Processing}, no.~2.\hskip 1em plus 0.5em minus 0.4em\relax IEEE, 2010, pp. 2606--2609.

\bibitem{jiodoa}
L.~{Wang}, R.~C. {de Lamare}, and M.~{Haardt}, ``Direction finding algorithms based on joint iterative subspace optimization,'' \emph{IEEE Transactions on Aerospace and Electronic Systems}, vol.~50, no.~4, pp. 2541--2553, October 2014.

\bibitem{mskaesprit}
S.~F.~B. Pinto and R.~C. de~Lamare, ``Multistep knowledge-aided iterative esprit: Design and analysis,'' \emph{IEEE Transactions on Aerospace and Electronic Systems}, vol.~54, no.~5, pp. 2189--2201, 2018.

\bibitem{Liu2016}
C.-L. Liu and P.~P. Vaidyanathan, ``{Super Nested Arrays: Linear Sparse Arrays With Reduced Mutual Coupling—Part I: Fundamentals},'' \emph{IEEE Transactions on Signal Processing}, vol.~64, no.~15, pp. 3997--4012, aug 2016.

\bibitem{Liu2016-2}
\BIBentryALTinterwordspacing
C.-l. Liu and P.~P. Vaidyanathan, ``{Super nested arrays: Sparse arrays with less mutual coupling than nested arrays},'' in \emph{2016 IEEE International Conference on Acoustics, Speech and Signal Processing (ICASSP)}.\hskip 1em plus 0.5em minus 0.4em\relax IEEE, mar 2016, pp. 2976--2980. [Online]. Available: \url{http://ieeexplore.ieee.org/document/7472223/}
\BIBentrySTDinterwordspacing

\bibitem{Leite2021}
W.~S. Leite and R.~C. de~Lamare, ``{List-Based OMP and an Enhanced Model for DOA Estimation With Nonuniform Arrays},'' \emph{IEEE Transactions on Aerospace and Electronic Systems}, vol.~57, no.~6, pp. 4457--4464, 2021.

\bibitem{Pal2010-1}
P.~Pal and P.~P. Vaidyanathan, ``{Nested arrays: A novel approach to array processing with enhanced degrees of freedom},'' \emph{IEEE Transactions on Signal Processing}, vol.~58, no.~8, pp. 4167--4181, 2010.

\bibitem{Pal2012}
------, ``{Nested arrays in two dimensions, Part I: Geometrical Considerations},'' \emph{IEEE Transactions on Signal Processing}, vol.~60, no.~9, pp. 4694--4705, 2012.

\bibitem{jioalrd}
L.~{Qiu}, Y.~{Cai}, R.~C. {de Lamare}, and M.~{Zhao}, ``Reduced-rank doa estimation algorithms based on alternating low-rank decomposition,'' \emph{IEEE Signal Processing Letters}, vol.~23, no.~5, pp. 565--569, May 2016.

\bibitem{Qiu2016}
\BIBentryALTinterwordspacing
L.~Qiu, Y.~Cai, R.~C. de~Lamare, and M.~Zhao, ``{Reduced-Rank DOA Estimation Algorithms Based on Alternating Low-Rank Decomposition},'' \emph{IEEE Signal Processing Letters}, vol.~23, no.~5, pp. 565--569, may 2016. [Online]. Available: \url{http://ieeexplore.ieee.org/document/7433376/}
\BIBentrySTDinterwordspacing

\bibitem{rdstapcp}
X.~Wang, Z.~Yang, J.~Huang, and R.~C. de~Lamare, ``Robust two-stage reduced-dimension sparsity-aware stap for airborne radar with coprime arrays,'' \emph{IEEE Transactions on Signal Processing}, vol.~68, pp. 81--96, 2020.

\bibitem{misc}
W.~Shi, Y.~Li, and R.~C. de~Lamare, ``Novel sparse array design based on the maximum inter-element spacing criterion,'' \emph{IEEE Signal Processing Letters}, vol.~29, pp. 1754--1758, 2022.

\bibitem{emisc}
X.~Sheng, D.~Lu, Y.~Li, and R.~C. de~Lamare, ``Enhanced misc-based sparse array with high udofs and low mutual coupling,'' \emph{IEEE Transactions on Circuits and Systems II: Express Briefs}, vol.~71, no.~2, pp. 972--976, 2024.

\bibitem{jidf}
R.~C. de~Lamare and R.~Sampaio-Neto, ``Adaptive reduced-rank processing based on joint and iterative interpolation, decimation, and filtering,'' \emph{IEEE Transactions on Signal Processing}, vol.~57, no.~7, pp. 2503--2514, 2009.

\bibitem{sjidf}
R.~Fa, R.~C. de~Lamare, and L.~Wang, ``Reduced-rank stap schemes for airborne radar based on switched joint interpolation, decimation and filtering algorithm,'' \emph{IEEE Transactions on Signal Processing}, vol.~58, no.~8, pp. 4182--4194, 2010.

\bibitem{aifir}
R.~de~Lamare and R.~Sampaio-Neto, ``Adaptive reduced-rank mmse filtering with interpolated fir filters and adaptive interpolators,'' \emph{IEEE Signal Processing Letters}, vol.~12, no.~3, pp. 177--180, 2005.

\bibitem{rrber}
Y.~Cai, R.~C. de~Lamare, B.~Champagne, B.~Qin, and M.~Zhao, ``Adaptive reduced-rank receive processing based on minimum symbol-error-rate criterion for large-scale multiple-antenna systems,'' \emph{IEEE Transactions on Communications}, vol.~63, no.~11, pp. 4185--4201, 2015.

\bibitem{jio}
R.~C. de~Lamare and R.~Sampaio-Neto, ``Reduced-rank adaptive filtering based on joint iterative optimization of adaptive filters,'' \emph{IEEE Signal Processing Letters}, vol.~14, no.~12, pp. 980--983, 2007.

\bibitem{jiocdma}
------, ``Reduced-rank space–time adaptive interference suppression with joint iterative least squares algorithms for spread-spectrum systems,'' \emph{IEEE Transactions on Vehicular Technology}, vol.~59, no.~3, pp. 1217--1228, 2010.

\bibitem{jiomimo}
------, ``Adaptive reduced-rank equalization algorithms based on alternating optimization design techniques for mimo systems,'' \emph{IEEE Transactions on Vehicular Technology}, vol.~60, no.~6, pp. 2482--2494, 2011.

\bibitem{jiostap}
R.~Fa and R.~C. De~Lamare, ``Reduced-rank stap algorithms using joint iterative optimization of filters,'' \emph{IEEE Transactions on Aerospace and Electronic Systems}, vol.~47, no.~3, pp. 1668--1684, 2011.

\bibitem{jiobf}
\BIBentryALTinterwordspacing
R.~de~Lamare, ``\BIBforeignlanguage{English}{Adaptive reduced-rank lcmv beamforming algorithms based on joint iterative optimisation of filters},'' \emph{\BIBforeignlanguage{English}{Electronics Letters}}, vol.~44, pp. 565--567(2), April 2008. [Online]. Available: \url{https://digital-library.theiet.org/content/journals/10.1049/el_20080627}
\BIBentrySTDinterwordspacing

\bibitem{wlmwf}
N.~Song, R.~C. de~Lamare, M.~Haardt, and M.~Wolf, ``Adaptive widely linear reduced-rank interference suppression based on the multistage wiener filter,'' \emph{IEEE Transactions on Signal Processing}, vol.~60, no.~8, pp. 4003--4016, 2012.

\bibitem{wljio}
N.~Song, W.~U. Alokozai, R.~C. de~Lamare, and M.~Haardt, ``Adaptive widely linear reduced-rank beamforming based on joint iterative optimization,'' \emph{IEEE Signal Processing Letters}, vol.~21, no.~3, pp. 265--269, 2014.

\bibitem{wlbd}
W.~Zhang, R.~C. de~Lamare, C.~Pan, M.~Chen, J.~Dai, B.~Wu, and X.~Bao, ``Widely linear precoding for large-scale mimo with iqi: Algorithms and performance analysis,'' \emph{IEEE Transactions on Wireless Communications}, vol.~16, no.~5, pp. 3298--3312, 2017.

\bibitem{l1stap}
Z.~Yang, R.~C. de~Lamare, and X.~Li, ``$l_1$ -regularized stap algorithms with a generalized sidelobe canceler architecture for airborne radar,'' \emph{IEEE Transactions on Signal Processing}, vol.~60, no.~2, pp. 674--686, 2012.

\bibitem{ccg}
\BIBentryALTinterwordspacing
L.~Wang, ``\BIBforeignlanguage{English}{Constrained adaptive filtering algorithms based on conjugate gradient techniques for beamforming},'' \emph{\BIBforeignlanguage{English}{IET Signal Processing}}, vol.~4, pp. 686--697(11), December 2010. [Online]. Available: \url{https://digital-library.theiet.org/content/journals/10.1049/iet-spr.2009.0243}
\BIBentrySTDinterwordspacing

\bibitem{sacg}
\BIBentryALTinterwordspacing
Z.~Yang, ``\BIBforeignlanguage{English}{Sparsity-aware space–time adaptive processing algorithms with l1-norm regularisation for airborne radar},'' \emph{\BIBforeignlanguage{English}{IET Signal Processing}}, vol.~6, pp. 413--423(10), July 2012. [Online]. Available: \url{https://digital-library.theiet.org/content/journals/10.1049/iet-spr.2011.0254}
\BIBentrySTDinterwordspacing

\bibitem{sastap}
Y.~Zhaocheng, R.~C. de~Lamare, and W.~Liu, ``Sparsity-based stap using alternating direction method with gain/phase errors,'' \emph{IEEE Transactions on Aerospace and Electronic Systems}, vol.~53, no.~6, pp. 2756--2768, 2017.

\bibitem{rdrcb}
S.~D. Somasundaram, N.~H. Parsons, P.~Li, and R.~C. de~Lamare, ``Reduced-dimension robust capon beamforming using krylov-subspace techniques,'' \emph{IEEE Transactions on Aerospace and Electronic Systems}, vol.~51, no.~1, pp. 270--289, 2015.

\bibitem{saalt}
R.~C. de~Lamare and R.~Sampaio-Neto, ``Sparsity-aware adaptive algorithms based on alternating optimization and shrinkage,'' \emph{IEEE Signal Processing Letters}, vol.~21, no.~2, pp. 225--229, 2014.

\bibitem{locsme}
H.~Ruan and R.~C. de~Lamare, ``Robust adaptive beamforming using a low-complexity shrinkage-based mismatch estimation algorithm,'' \emph{IEEE Signal Processing Letters}, vol.~21, no.~1, pp. 60--64, 2014.

\bibitem{okspme}
------, ``Robust adaptive beamforming based on low-rank and cross-correlation techniques,'' \emph{IEEE Transactions on Signal Processing}, vol.~64, no.~15, pp. 3919--3932, 2016.

\bibitem{lrcc}
------, ``Distributed robust beamforming based on low-rank and cross-correlation techniques: Design and analysis,'' \emph{IEEE Transactions on Signal Processing}, vol.~67, no.~24, pp. 6411--6423, 2019.

\bibitem{Fortunati2014}
S.~Fortunati, R.~Grasso, F.~Gini, M.~S. Greco, and K.~LePage, ``{Single-snapshot DOA estimation by using Compressed Sensing},'' \emph{EURASIP Journal on Advances in Signal Processing}, no.~1, p. 120, 2014.

\bibitem{Pati1993}
Y.~Pati, R.~Rezaiifar, and P.~Krishnaprasad, ``{Orthogonal matching pursuit: recursive function approximation with applications to wavelet decomposition},'' in \emph{Proceedings of 27th Asilomar Conference on Signals, Systems and Computers}, vol.~1.\hskip 1em plus 0.5em minus 0.4em\relax IEEE Comput. Soc. Press, 1993, pp. 40--44.

\bibitem{Blumensath2009}
\BIBentryALTinterwordspacing
T.~Blumensath and M.~E. Davies, ``{Iterative hard thresholding for compressed sensing},'' \emph{Applied and Computational Harmonic Analysis}, vol.~27, no.~3, pp. 265--274, 2009. [Online]. Available: \url{http://dx.doi.org/10.1016/j.acha.2009.04.002}
\BIBentrySTDinterwordspacing

\bibitem{dce}
S.~Xu, R.~C. de~Lamare, and H.~V. Poor, ``Distributed compressed estimation based on compressive sensing,'' \emph{IEEE Signal Processing Letters}, vol.~22, no.~9, pp. 1311--1315, 2015.

\bibitem{Gu2011}
\BIBentryALTinterwordspacing
J.-F. Gu, W.-P. Zhu, and M.~Swamy, ``{Minimum redundancy linear sparse subarrays for direction of arrival estimation without ambiguity},'' in \emph{2011 IEEE International Symposium of Circuits and Systems (ISCAS)}.\hskip 1em plus 0.5em minus 0.4em\relax IEEE, may 2011, pp. 390--393. [Online]. Available: \url{http://ieeexplore.ieee.org/document/5937584/}
\BIBentrySTDinterwordspacing

\bibitem{Elbir2019}
A.~M. Elbir, S.~Mulleti, R.~Cohen, R.~Fu, and Y.~C. Eldar, ``{Deep-Sparse Array Cognitive Radar},'' \emph{2019 13th International Conference on Sampling Theory and Applications, SampTA 2019}, pp. 1--4, 2019.

\bibitem{See2004}
C.~M.~S. See and A.~B. Gershman, ``{Direction-of-arrival estimation in partly calibrated subarray-based sensor arrays},'' \emph{IEEE Transactions on Signal Processing}, vol.~52, no.~2, pp. 329--338, feb 2004.

\bibitem{Swindlehurst2001}
A.~Swindlehurst, P.~Stoica, and M.~Jansson, ``{Exploiting arrays with multiple invariances using MUSIC and MODE},'' \emph{IEEE Transactions on Signal Processing}, vol.~49, no.~11, pp. 2511--2521, 2001.

\bibitem{Rieken2004}
\BIBentryALTinterwordspacing
D.~Rieken and D.~Fuhrmann, ``Generalizing music and mvdr for multiple noncoherent arrays,'' \emph{IEEE Transactions on Signal Processing}, vol.~52, pp. 2396--2406, 9 2004. [Online]. Available: \url{http://ieeexplore.ieee.org/document/1323249/}
\BIBentrySTDinterwordspacing

\bibitem{Tirer2021}
T.~Tirer and O.~Bialer, ``{Direction of arrival estimation for non-coherent sub-arrays via joint sparse and low-rank signal recovery},'' \emph{ICASSP, IEEE International Conference on Acoustics, Speech and Signal Processing - Proceedings}, vol. 2021-June, pp. 4395--4399, 2021.

\bibitem{cvx}
M.~Grant and S.~Boyd, ``{CVX}: Matlab software for disciplined convex programming, version 2.2,'' \url{http://cvxr.com/cvx}, Mar. 2020.

\bibitem{Moffet1968}
A.~Moffet, ``{Minimum-redundancy linear arrays},'' \emph{IEEE Transactions on Antennas and Propagation}, vol.~16, no.~2, pp. 172--175, mar 1968.

\bibitem{Tuncer:2009:CMD:1795749}
T.~E. Tuncer and B.~Friedlander, \emph{Classical and Modern Direction-of-Arrival Estimation}.\hskip 1em plus 0.5em minus 0.4em\relax Orlando, FL, USA: Academic Press, Inc., 2009.

\end{thebibliography}

\end{document}